 \documentclass[preprint, 12pt]{elsarticle}

 \usepackage{amssymb}

 \usepackage{amsmath}
 \newtheorem{theorem}{Theorem}

 \newtheorem{defi}{Definition}
 \newtheorem{example}{Example}
 \newtheorem{property}{Property}
 
 \journal{Journal Name}

\begin{document}
	
	 \begin{frontmatter}
		
		

	 \title{Relations among Open-loop Control Ability, Control Strategy Space and Closed-loop Performance for Linear Discrete-time Systems 
		 \thanks{Work supported by the National Natural Science Foundation of China (Grant No. 61273005)}}
		
		
		 \author{Mingwang Zhao}
		
		 \address{Information Science and Engineering School, Wuhan University of Science and Technology, Wuhan, Hubei, 430081, China \\
			Tel.: +86-27-68863897 \\
		Work supported by the National Natural Science Foundation of China (Grant No. 61273005)}
		
		 \begin{abstract}
			In this article, the definition on the control ability, and the relation between the open-loop control ability and the closed-loop performance are studied systematically for the linear dynamical systems. Firstly, to define and compare rationally the state control ability between the different controlled plants or one controlled plant with the different system parameters, the normalization of the input variables, the state variables, the system mdeols are discussed. With the help of the normalization, the state control ability with the time attribute can be difined under the unit input constraint (input amplitude limited). And then, a theorem on the relations among the open-loop control ability, the control strategy space (i.e., the solution space of the input variables for control problems), and the closed-loop performance with the time attribute is purposed and proven. Based on that, the conclusion that it is very necessary to optimize the control ability for the practical engineering problems can be got. Finally, the simulation experiments show us the normalizing the variables and system models, and comparing the contol ability between the different controlled palnts. 
			
		 \end{abstract}
		
		
	 \end{frontmatter}
	
	
	 \section{Introduction}
	 \label{S:1}

Putting forward the concept and criterion on the state controllability of the dynamical systems in 1960's by R. Kalman, et al, \cite{KalmHoNar1963} initiated a new era for control theory. As we known, the concept can reveal deeply the possibility controlling the state variables by the input variables and impels us to understand and control well the dynamical systems. Therefore, the concept became one of the most important concept to support the 60 years development of the control theory.
	
	It is a pity that the controllability concept is only a qualitative concept with two-value logic and the dynamical systems is distinguished as only two classes of systems, controllable systems and uncontrollable systems, according to the corresponding controllability criterion. The concept and criterion could not tell us the control ability and control efficiency of the input variables to the state variables, and the quantitative concept and analysis method on that are failed to establish. In fact, the quantitative concept and analysis method are very important for the control theory and engineering and many engineering problems are dying to these concept and method for getting the easier controller desgin process, the better controller, and the better closed-loop performance index. For example, evaluating the control ability and control efficiency of the controlled plants can help us to understand and solve the following important control problems: 
	
	1) how to choose the controlled plants or equipments (e.g., choosing DC or AC motor for designing in some electric drive system?), how to choose the input variables(e.g., choosing power supply of main circuit or excitation circuit as the input variable for desiging in some DC motor speed-controller), and how to place the location of the actuators in larger mechanical system (e, g., mechanical cantilever, bridge, solar panels, etc), for maximizing the control ability of the open-loop systems. 
	
	2) how to design and optimize the structure and technical parameters of the open-loop plants to get the more control ability and then to make designing and implementing the closed-loop controller easily.
	
	3) how to determine the leader, the sub-leaders, and the connections between the nodes in the networked control systems and formation system for maximizing the performance of these systems.
		
4) how to determine reasonably the expected target state or state trace, the control horizon and the optimization horizon for these optimal control problems, adaptive control problems, predictive control problems, and the receding-horizon control(RHC) problems, and then the control laws can be got by solving these control problems.

To summarize above, defining, quantifying and optimizing the control ability are with the very greater signification for the control theory and engineering.

In this paper, the definition on the control ability is studied systematically. Firstly, to define and compare rationally the state control ability of the input variables betweenthe different control plants or one controlled plant with the different system parameters, the normalization of the input variables, the state variables, and the system models are discussed. With the help of the normalization, the time-attribute control ability with the unit input variables can be defined. And then, a theorem on the relations among the open-loop control ability, the control strategy space (i.e., the solution space of the input variables for control problems), and the closed-loop time performance is purposed and proven. Based on that, the conclusion that it is necessary to optimize the control ability for the practical engineering problems can be got. Finally, the simulation experiments show us the normalizing the variables and system models, and comparing the time-attribute contol ability between the different controlled palnts. 
	 
 \section{Normalization and Constraints for the Variables and System Models}
	
	In this paper, the linear discrete-time (LDT) systems is as a sample for studying the definition and the analysis method on the state control ability, and the obtained results can be generalized conveniently to other classes of dynamical systems. In general, the LDT Systems can be formulated as follows:
 \begin{equation}
x_{k+1}=Ax_{k}+Bu_{k}, \quad x_{k} \in R^{n}, u_{k} \in R^{r}, \label{eq:a0201}
 \end{equation}
 \noindent where $x_{k}$ and $u_{k}$ are the state variables and input
variables, respectively, and matrices $A \in R^{n \times n}$ and $B \in
R^{n \times r}$ are the state matrix and input matrix, respectively, 
in the system models \cite{Kailath1980}, \cite{Chen1998}. 
To investigate the controllability of the
linear dynamic systems \eqref{eq:a0201}, the controllability matrix and the controllability Grammian matrix can be defined as follows
 \begin{align}
P_{N} & = \left[ B, AB, \dots, A^{N-1}B \right ]
 \label{eq:a0202} \\
G_{N} & = \sum _{i=0} ^{N-1} A^{i}B \left( A^{i}B \right)^T \label{eq:a0203}
 \end{align}	
where $N \ge n$. That the rank of the matrix $P_{N}$ and $G_{N}$ is $n$, that is, the dimension of the state space the systems \eqref{eq:a0201}, is the well-known sufficient and necessary criterions on the state controllability for the LDT systems. 

In papers \cite{VanCari1982}, \cite{Georges1995}, \cite{PasaZamEul2014}, and \cite{Ilkturk2015}, the determinant value $ \det \left( G_{N} \right) $ and the minimum eigenvalue $ \lambda _{ \textnormal {min} } \left (G_{N} \right ) $ of the controllability Grammian matrix $G_{N}$ can be used to quantify the control ability of the input variables to the state space, and then be chosen as the objective function for optimizing and promoting the control ability of the linear dynamical systems. Due to lack of the analytical computing of the determinant $ \det \left( G_{N} \right) $ and eigenvalue $ \lambda _{ \textnormal {min}} \left (G_{N} \right ) $, these optimizing problems for the control ability are solved very difficulty, and few achievements about that were made. Out of the need of the practical control engineering, quantifying and optimizing the control ability are key problems in control theory and engineering fields.

To study rationally the control ability of the input variables to the state variables in different dynamical systems, it is necessary to normalize the input variables, the state variables, and the system models. Based on the normalization, the control ability can defined and discussed in detail.
	
 \subsection{Normalization of the Variables and System Models}
	
In different practical controlled plants, the physical dimensions, scales, value ranges of the input variables and state variables are different. Comparing rationally the control ability of these different practical plants, or these different input variables, firstly, the input variables, the state variables, and then the system models must be normalized according to the practical control problems. For example, to compare the control ability between the two different input variables in one controlled plant or two different controlled plants, the physical dimensions, scales, value ranges are necessary to be adjusted as a proper compatible values with some rationalness. Similarly, to compare the controlled ability between the two different state variables, the dimensions, scales, ranges are also necessary to be adjusted as a proper compatible values. Next, two examples are discussed for showing these adjustment and normalization.

1) If only one input variable can be used to be designed the speed controller of a practical DC motor, which voltage variable, the power supply of the main circuit or excitation circuit, is chosen as that for maximizing control ability? The value ranges of these voltage variables and the ratios between the voltage variables and the speed variables of the motor must be adjusted to be with uniformity. Based on this, comparing with the different input variables is with rationalness and signification.

For example, if the rated values of the main circuit input variable $u_m(k)$ and excitation circuit input variable $u_e(k)$ are respectively $u_m^*$ and $u_e^*$, that is, the input variables are respectively in the rated interval $[-u_m^*, u_m^*]$ and $[-u_e^*, u_e^*]$, to compare rational the control ability of the input variables, the input variables of the system models for the two cases should be normalized respectively as 
 \begin{align} u_k=u_m(k)/u_m^* \; \textnormal{ and } \; u_k=u_e(k)/u_e^* \label{eq;a0304}
 \end{align}
where $u_k$ is the normalized input variable which the define domain is $[-1, 1]$. If the system models with the two input variables are respectively $ \Sigma \left( A_m, B_m \right)$ and $ \Sigma \left( A_e, B_e \right)$, the normalized system models are respectively as 
 \begin{align} \Sigma \left( A_m, u_m^*B_m \right) \; \; \textnormal {and} \; \; \Sigma \left( A_e, u_e^*B_e \right) \label{eq;a0305}
 \end{align}
2) Which motor, DC motor with the excitation controller or AC motor with the variable frequency controller, can be determined to be used to the some electric speed control system for maximizing control ability? 
The value ranges of the input variables and the state variables, the ratios between the input variables and the speed variable, and the power of the electric energy of the two motors, must be adjusted to be with uniformity. Based on this, comparing with the different controlled plants is with rationalness and signification.

For example, let the rated values of the inputs $u_d(k)$ and $u_a(k)$, the main circuit currents $i_d(k)$ and $i_a(k)$, the speed outputs $y_d(k)$ and $y_a(k)$, and the accelerations $ \dot y_d(k)$ and $ \dot y_a(k)$ of the two motors are respectively as
 \begin{align} \left( u_d^*, \; i_d^*, \; y_d^*, \; \dot y_d^* \right) \; \; \textnormal {and} \; \;
 \left( u_a^*, \; i_a^*, \; y_a^*, \; \dot y_a^* \right) \notag
 \end{align}

For comparing the control ability of the two motors when the all variables in there speed rated intervals, the input and output variables should be normalized as 
 \begin{align} 
 \left \{ \begin{array}{l} \left( u_d(k)/u_d^*, \; i_d(k)/i_d^*, \; y_d(k)/y_d^*, \; \dot y_d(k)/ \dot y_d^* \right) \\
 \left( u_a(k)/u_a^*, \; i_a(k)/i_a^*, \; y_a(k)/y_a^*, \; \dot y_a(k)/ \dot y_d^* \right) \end{array} 
 \right. \label{eq:a0206}
 \end{align}
And then, the all input and output variables are in the define domain $[-1, 1]$ and the system models $ \Sigma \left( A_d, B_d \right)$ and $ \Sigma \left( A_a, B_a \right)$ with the state variable vectors $ \left[i_d, y_d, \dot y_d \right ] $ and $ \left[i_a, y_a, \dot y_a \right ] $ and are transformed respectively as 
 \begin{align} \label{eq:a0207}
 \Sigma \left( P_d^{-1} A_dP_d, u_d^*P_d^{-1}B_d \right) \; \; \textnormal {and} \; \; \Sigma \left(P_a^{-1} A_aP_a, u_a^*P_a^{-1}B_a \right)
 \end{align}
where the normalization matrices $P_d$ and $P_a$ for the state variables are respectively as
 \begin{align}
P_d= \textnormal {diag} \left \{i_d^*, y_d^*, \dot y_d^* \right \} \; \; \textnormal {and} \; \;
P_a= \textnormal {diag} \left \{i_a^*, y_a^*, \dot y_a^* \right \} \label{eq:a0208}
 \end{align}

For comparing the control ability of the two motors which the all output variables are in the same expecting interval as follows
 $$[-i_s, i_s], \;[-y_s, y_s], \;[- \dot y_s, \dot y_s], \;$$
 the input and output variables should be normalized as 
 \begin{align} 
 \left \{ \begin{array}{l} \left( u_d(k)/u_d^*, \; i_d(k)/i_s, \; y_d(k)/y_s, \; \dot y_d(k)/ \dot y_s \right) \\
 \left( u_a(k)/u_a^*, \; i_a(k)/i_s, \; y_a(k)/y_s, \; \dot y_a(k)/ \dot y_s \right) \end{array} \label{eq:a0209}
 \right.
 \end{align}
And then, the system models are transformed respectively as 
 \begin{align} \label{eq:a0210}
 \Sigma \left( P_s^{-1} A_dP_s, u_d^*P_s^{-1}B_d \right) \; \; \textnormal {and} \; \; \Sigma \left(P_s^{-1} A_aP_s, u_a^*P_s^{-1}B_a \right)
 \end{align}
where the normalization matrices $P_s$ is as
 \begin{align} \label{eq:a0211}
P_s= \textnormal {diag} \left \{i_s^*, y_s^*, \dot y_s^* \right \} 
 \end{align}

After the normalization of the system models as above, analyzing, comparing and optimizing the control ability between the different dynamical systems are with rationality. The above normliztion methods are also applied to other controlled plants.

 \subsection{Constraints of the amplitude, fule and Energy of the Input Variables }

The so-called state control ability is indeed the ability controlling the state variables by the input variables. The basis for comparison is the normalization and constraint conditions of the input variables. In fact, the input variables of the most practical controlled plants are bounded, with some constraints, or some with saturation element \cite{Bernstein1995}, \cite{Hulin2001}, \cite{Hu2002}. For example, the power supply voltage variables as the input variables for the DC or AC motor are bounded, and the total fuel or energy wasted in the rockets is with some constraints. Therefore, based on these bounded values and constraints, the input variables can be normalized.

In control theory and engineering field, the most common bounded and constraint cases of the input variable vector $u_k$ can be summarized as the two following cases
	 \begin{align} U_{a, p} &= \left \{ u_k: \Vert u_k \Vert_p = \left ( \sum _{i=0} ^{r} \vert u_{k, i} \vert ^p \right ) ^{1/p} \le U \right \} \label{eq:a0212} \\
	U_{t, p} &= \left \{ U_N: \Vert U_N \Vert_p = \left ( \sum _{k=0} ^{N-1} \Vert u_k \Vert _p ^p \right ) ^{1/p}
 \le U \right \} \label{eq:a0213}
 \end{align} 
where $r$ and $u_{k, i}$ are the input variable numbers and the $i$-th input variable of the multi-input systems, respectively; $U_N= \left[u_0^T, u_1^T, \dots, u_{N-1}^T \right]^T $. The constraints with $p=1, 2, 3$ are respectively the amplitude, fuel, and energy bounded. The condition \eqref{eq:a0212} is for bounding the input variables in the all sampling times, and the condition \eqref{eq:a0213} is for constraining the total waste of the input in a control period $[1, N]$. In fact, for the single-input systems, the constraints \eqref{eq:a0212} are a same constraint as 
	 \begin{align} \vert u_k \vert \le U, \; \forall k \ge 0 \label{eq:a0214} 
 \end{align} 

In practical control engineering problems, the most common constrints are as follows
	 \begin{align} \Vert u_k \Vert_ \infty & \le U, \; \forall k \ge 0 \label{eq:a0215} \\
 \Vert U_N \Vert_1 & \le U \label{eq:a0216} \\
 \Vert U_N \Vert_2 & \le U \label{eq:a0217}
 \end{align} 
These 3 constraints are bounded on the amplitude, total fuel, and total energy of the input variables, respectively. For comparing conveniently the control ability, the bounded value $U$ is chosen as 1, and then these 3 constraints can be called as the unit input constraint, unit total fuel constraint, and unit total energy constraint.
 
 \section{The Definitions of the Controllability Region }
	
Similar to the definition of the state controllability region for the input-saturated linear systems in papers \cite{Bernstein1995}, \cite{Hulin2001}, \cite{Hu2002}, the state controllability regions of the LDT Systems with the input constraints are defined as follows
	 \begin{align} 
	R^c_{*}(N) & = \left \{x : x=- \sum _{k=0}^{N-1} A^{-k-1 } Bu_k, \; \; \forall u_k \in U_* \right \} \notag \\ 
	& = \left \{A^{-1}x : x= \sum _{k=0}^{N-1} A^{-k } Bz_k, \; \; \forall z_k \in U_* \right \} \label{eq:a0218} \\
		R^d_{*}(N) & = \left \{x : x= \sum _{k=0}^{N-1} A^{N-k-1 } Bu_k, \; \; \forall u_k \in U_* \right \} \notag \\ 
		& = \left \{x : x= \sum _{k=0}^{N-1} A^{k } Bz_k, \; \; \forall z_k \in U_* \right \} \label{eq:a0219}
	 \end{align}
where * indicate the case of the input constraints in Eqs. \eqref{eq:a0212} and \eqref{eq:a0213}, $R^c_{*}(N)$ and $R^d_{*}(N)$ are the narrow controllability regions ( a.k.a. "recover region") and the reachability region, respectively. 	By Eqs. \eqref{eq:a0218} and \eqref {eq:a0219}, we can see, the narrow controllability regions and the reachability region can be transformed each other as follows
 \begin{align}
 \left. R^c_{*} \right \vert _A = A^{-1} * \left. R^d_{*} \right \vert _{A^{-1}} \label{eq:a0220}
 \end{align}
Therefore, the broad controllability region, include the narrow controllability regions and the reachability region, can be defined as follows
	 \begin{align} 
R_{*}(N) 
 = \left \{x : x= \sum _{k=0}^{N-1} A^{k } Bz_k, \; \; \forall z_k \in U_* \right \} \label{eq:a0221} 
 \end{align}
In fact, the reahability region $R^d_{*} (N)$ is identical with the broad controllability region $R_{*}(N)$, and the narrow controllability region $R^c_{*}(N)$ can be got by transformed of $R_{*}(N)$. Next, the control ability for the broad controllability region $R_{*}(N)$ is discused and the obtained results can be generalized to the regions $R^c_{*}(N)$ and $R^d_{*}(N)$.

For the 3 most common constraints in Eqs. \eqref{eq:a0215}, \eqref{eq:a0216}, and \eqref{eq:a0217}, the controllability regions, i.e., $ R_{a, \infty}(N)$, $R_{t, 1}(N)$, and $R_{t, 2}(N)$, are the biggest range of the controllable state with the unit input constraint, unit total fuel constraint, and unit total energy constraint, respectively. 

The state controllability region $R_{*}(N) $ are as a convex geometry in $n$-dimensional (abbreviation: $n$-D) space. Region $ R_{a, \infty}(N)$ is a parallel polyhedron and can be regarded as a special zonotope \cite{Hulin2001}, \cite{Hu2002}, \cite{zhaomw202001}, region $ R_{t, 2}(N)$
is a ellipsoid (i.e., so called "controllability ellipsoid" ) \cite{DulPag2000}, \cite{KURVARA2007}, \cite{PolNaKhl2008}, \cite{nkh2016}, \cite{cannkh2017}, but the region $ R_{t, 1}(N)$ is a rhomboid. In the following sections, the controllability region $R_{a, \infty}(N) $ with the unit input variable constraint $U_{a, \infty}$ and the corresponding control ability are defined and analyzed, and other cases, the controllability regions $ R_{t, 2}(N)$ and $ R_{t, 1}(N)$, will be studied in another paper. 

 \section{Control Ability under the Unit Input Constraint}
	
 \subsection {The properties of the controllability Region $R_{a, \infty}(N) $ }
	
As stated in papers \cite{Hulin2001}, \cite{Hu2002}, \cite{zhaomw202001}, the state controllability region $R_{a, \infty}(N) $ (Short as $R(N)$ )under the unit input constraint is a convex geometry, can be regard as a parallel polyhedron or a special zonotope. In fact, the region $R(N) $ is surrounded by a series of vertices, edges, 2-dimensinal faces, 3-D feces, \dots .
All vertices, edges, $i$-D faces ($i= \overline{2, n-1}$) construct the boundary of the region $R(N) $.

Some properties about the vertices, shape and size of the region can be summarized as follows \cite{Hulin2001}, \cite{Hu2002}, \cite{zhaomw202001}.

 \begin{property} \label{pr:p0301}
The all vertice of the controllability region $R(N) $ can be computed as follows 
 \begin{align}
 \textnormal {Ver} \left ( R(N) \right ) =
 \left \{ x \left|x= \sum_{i=0}^{N-1} \textnormal{sgn} \left(d^{T} A^{i}B \right)A^{i}B, \; \forall d \in R^{n} \right. \right \} \label{eq:a0222}
 \end{align}
 \end{property}
	
	Based the vertice produced by Eq. \eqref{eq:a0222}, all edges and $i$-dimensinal faces ($i= \overline{2, n-1}$) can be produced recurssively.

 \begin{property} \label{pr:p0302}
If the LDT Systems \eqref{eq:a0201} is controllable, when $N \ge n$ the controllability region $R(N) $ is a $n$-D geometry, and then for any $N_1<N_2$, we have
 \begin{align} 
	R (N_1) \subset R(N_2) \; \textnormal {and} \; \partial R (N_1) \cap \partial R(N_2) = \phi \label{eq:a0223} 
			 \end{align}
that is, the geometry $R$ is strictly monotonic expansion, where $ \partial R $ is the boundary of the geometry $R$. 

If the systems is not controllable, when $N \ge n$ the region $R(N) $ is a $n_c$-D geometry, and then for any $N_1<N_2$, we have
		 \begin{align} 
			R (N_1) \subseteq R(N_2) \; \textnormal {and} \; \partial R (N_1) \cap \partial R(N_2) \neq \phi \label{eq:a0224} 
				 \end{align}
that is, the geometry $R$ is monotonic expansion, where $n_c$ is the controllability index, that is, $n_c= \textnormal {rank} P_N$. 
	 \end{property}

\begin{figure}[htbp]
	 \centering
	 \includegraphics[width=0.7 \textwidth]{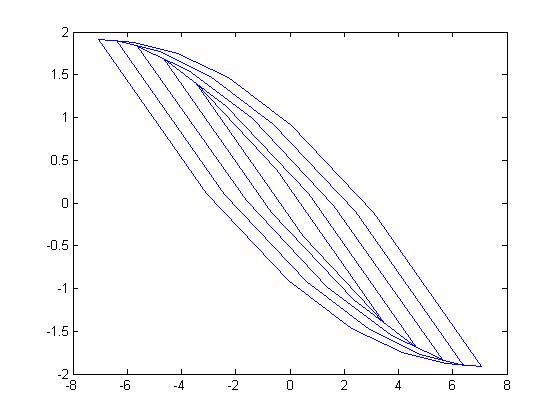} 
	 \caption[c]{The 2-D zonotopes $R^c(N)$ when $N \in \{2, 6 \}$ \label {fig:aa03}}	
 \end{figure}

 \textbf{Fig. \ref{fig:aa03}} shows the 2-D narrow controllability region $R ^c (N)$ generated by the matrix pair $(A, B)$ as follows
$$
A= \left[ \begin{array}{cc}
 1.1616 & -0.5051 \\
-0.0505 & 1.6162
 \end{array} \right], \; \; 
b= \left[ \begin{array}{c}
1.8182 \\
-0.8182
 \end{array} \right]
$$

Because that the matrix pair $(A, B)$ is controllable, the controllability region $R^c(N)$ in \textbf{Fig. \ref{fig:aa03}} is strictly monotonic expansion
with the increase of the sampling step $N$.

In the next discussion, the systems are assumed always as a controllable systems and the geometry $R$ is a $n$-D zonotope.

 \subsection {The Definition of the Control Ability }
	
As we know, the bigger of the controllability region $R$ is, the more the controllable states in the state space are, and then, can we say that the control ability of the dynamical systems is the stronger or not.
In the next subsection, it will proven that for the control problem stabilizing a given initial state to the origin of the state space (or making the initial state at the origin to reach the given state ), 
the bigger of the controllability region, the bigger the solution space of the input variables is, and then the better the some closed-loop control performance is. Thus, the size of the controllability region $R$ is used to define and describe the control ability.

Next, two equivalent definitions on the stronger control ability between the two differential controlled plants, or of one controlled plants with two sets of systems parameters are purposed as follows.

 \begin{defi} \label {de:d0301}
		For the controllability regions $R^{(1)}(N)$ and $ R^{(2)}(N)$ of the two given LDT Systems $ \Sigma_1$ and $ \Sigma_2$, 
		if the given $x_0 \in \partial R^{(1)}(N_1) \cap \partial R^{(2)}(N_2)$ and $N_1 < N_2$, the time-attribute control ability of the systems $ \Sigma_1$ at state $x_0$ is stronger than the systems $ \Sigma_2$. If for any $N>0$, the time-attribute control ability of the systems $ \Sigma_1$ at all state in $ \partial R^{(1)}(N) \cap \partial R^{(2)}(N)$ 
		is stronger than the systems $ \Sigma_1$, the control ability of the systems $ \Sigma_1$ is stronger than the systems $ \Sigma_2$.
	 \end{defi}

 \begin{defi} \label {de:d0302}
For any $N>0$, if the two controllability regions $R^{(1)}(N)$ and $ R^{(2)}(N)$ of the LDT Systems $ \Sigma_1$ and $ \Sigma_2$ satisfy 
 \begin{align} 
R^{(1)} (N) \supset R^{(2)}(N) \; \textnormal {and} \; \partial R^{(1)} (N) \cap \partial R^{(2)}(N) = \phi \label{eq:a0225}, 
 \end{align}
the time-attribute  control ability of the systems $ \Sigma_1$ is stronger than the systems $ \Sigma_2$. if $R^{(1)}(N)$ and $ R^{(2)}(N)$ for any $N$ satisfy 
 \begin{align} 
	R^{(1)} (N) \supseteq R^{(2)}(N) \; \textnormal {and} \; \partial R^{(1)} (N) \cap \partial R^{(2)}(N) \neq \phi \label{eq:a0226}, 
 \end{align}
the time-attribute control ability of the systems $ \Sigma_1$ is not weaker than the systems $ \Sigma_2$.
 \end{defi}

According to the above definitions, the bigger the size the controllability region $R(N)$ is, and the stronger the time-attribute control ability of the LDT systems is. Therefore, based on the computing and analyzing the size and shape of the geometrys $R(N)$, the time-attribute control ability of the dynamical systems can be compared for many control engineering problems as stated above. Except for the time-attribute control ability, based on  other input constraints, the energe-attribute and fuel-attribute control ability can be defined and analyzed, and the related works will be carried out in another papers. 

 \section{Theorem on Relation between the Open-loop Control Ability and the Closed-loop Performances}

In this section the control ability for the unit input constraint is discussed in detail and the results can be generalized convenient to the other constraints of the input variables. Before the discussion, a time-optimal property for the boundary of the controllabilty region is stated as follows \cite{Hulin2001}, \cite{Hu2002}.

 \begin{property}
	 \label{pr:p0303}
	If the LDT Systems \eqref{eq:a0201} is controllable (or reachable) and the given state $x_0$ satisfy
	 \begin{align} 
	 x_{0} \in R (N) \setminus R (N-1) \label{eq:a0227}
	 \end{align}
the time waste of the time-optimal control problem for stabilizing the state $x_0$ to the origin of the state space (or making the state at the origin to reach the given state $x_0$ ) under the input amplitude constraint, is $N$, that is, the fewest control sampling number is $N$.
 \end{property}

Based on the definition of the time-attribute control ability and above properties, a theorem on the
 relations among the open-loop control ability, the solution space of the input variables, and the closed-loop performances are purposed and proven as follows.

 \begin{theorem} \label{th:t0301} 
It is assumed that two LDT Systems $ \Sigma_1$ and $ \Sigma_2$ are controllable (or reachable), and their controllability regions are $R^{(1)} (N) $ and $R^{(2)} (N)$ respectively.
If we have
	 \begin{align} \label{eq:a0228}
	R^{(1)} (i) \subseteq R^{(2)} (i), \; \forall i \le N, 
	 \end{align}
for the control problem stabilizing the state $x_0 \left (x_0 \in R^{(1)} (N) \cap R ^{(2)} (N) \right) $ to the origin of the state space (or making the state at the origin to reach the given state $x_0$ ), 
the following conclusions hold under the input amplitude constraint.

1) The time waste of the time-optimal control for the system $ \Sigma_2$ is not more than that of $ \Sigma_1$, that is, there exist some control strategies with the less control time and the faster response speed for the system $ \Sigma_2$. 

2) There exist more control strategies for the system $ \Sigma_2$, that is, 
 the bigger the controllability region is, the bigger the solution space of the input variables for the control problems, and then the easier designing and implementing the controller are. 
 \end{theorem}

{ \bfseries Proof of Theorem \ref{th:t0301} } Next, the proof will be discussed only for the stabilizing control problem, and the obtained result holds for the reaching control problem.

First, according to the definition of the controllability region, for the state controllable systems $ \Sigma_1$ and $ \Sigma_2$, we have, 
 \begin{align} \label{eq:a0229}
R^{(j)} (i) \subset R^{(j)} (i), \; j=1, 2; \;i=1, 2, \dots , N-1
 \end{align}
So, by Eq. \eqref{eq:a0228} and Eq. \eqref{eq:a0229}, we know, for any state $x_{0} \in R^{(1)} (N) \setminus R^{(1)} (1)$, there must exist two finite positive number $k_1$ and $k_2(k_2 \le k_1 \le N)$ satisfied
 \begin{align} 
& x_{0} \in \left \{ R^{(1)} \left(k_1 \right) \setminus R ^{(1)} \left (k_1-1 \right) \right \} \cap R ^{(2)} \left (k_2 \right) \label{eq:a0230} \\
& R^{(1)} \left(k_1 \right) \subseteq R ^{(2)} \left( k_2 \right) \label{eq:a0231}
 \end{align}

Therefore, for controlling the system state variables from the given $x_0$ to the origin, the fewest sampling steps must be $k_1$ for the system $ \Sigma_1$, but must be less than or equal to $k_2$ for the system $ \Sigma_2$. So, for the any state $x_{0} \in R^{(1)} (N) \setminus R^{(1)} (1)$, for $k_2 \le k_1$, the time waste of the time-optimal control for the system $ \Sigma _2$ is not more than that of the system $ \Sigma_1$.

In addition, for any state $x_{0} \in R^{(1)} (1)$, the fewest control times (sampling steps) are 1 for both of the two systems, that is, the time waste of the time-optimal control for the system $ \Sigma _2$ is not more than that of the system $ \Sigma_1$.

In summary, for any state $x_{0} \in R^{(1)} (N) \cap R^{(2)} (N)$, the conclusion 1) in { \bfseries Theorem \ref{th:t0301} } holds.

(2) Denoting the input sequence and its solution space for controlling the given state $x_{0}$ to the origin for systems $ \Sigma _i$ as $u^{(i)}_{0, N-1}(x_{0})$ and $U^{(i)}_{N}(x_{0})$, respectively. Without loss of the generality, it is assumed that $r=1$, that is, the systems are single-input systems. So, by Eq. \eqref{eq:a0228} and Eq. \eqref{eq:a0229}, we know, for any state $x_{0} \in R^{(1)} (N)$, there must exist two finite positive number $k_1$ and $k_2(k_2 < k_1 \le N)$ satisfied one of the following conditions
 \begin{align} 
(a1) \; & x_{0} \in  \partial R^{(1)} \left(k \right)    \cap  \partial R ^{(2)} \left(k \right), \;\;k=1,k_1 \label{eq:a0232a} 
 \\
(a2) \; & x_{0} \in  \partial R^{(1)} \left(k \right)    \cap   \widetilde R ^{(2)} \left(k \right), \;\;k=1,k_1 \label{eq:a0232b} 
\\
(a3) \; & x_{0} \in  \widetilde R^{(1)} \left(k \right)    \cap  \widetilde R ^{(2)} \left(k \right), \;\;k=1,k_1 \label{eq:a0232c} 
\\
(b1) \; & x_{0} \in  \partial R^{(1)} \left(k_1 \right)    \cap  \partial R ^{(2)} \left(k_2 \right)  \label{eq:a0232d} 
\\
(b2) \; & x_{0} \in  \partial R^{(1)} \left(k_1 \right)    \cap   \widetilde R ^{(2)} \left(k_2 \right)  \label{eq:a0232e} 
\\
(b3) \; & x_{0} \in \widetilde R^{(1)} \left(k_1 \right)    \cap  \widetilde R ^{(2)} \left(k_2 \right)  \label{eq:a0232f} 
 \end{align}
 where $ \widetilde R(k)=\left (R (k) \setminus \partial R(k) \right) \setminus R(k-1)$. For controlling the state $x_0$ to the origin, the input sequence $u^{(i)}_{0, N-1}(x_{0})$ must satisfies the following state eauation.
  \begin{align} \label{eq:a0236} 
 x_0= \left[ A^{-1}B, A^{-2}B, \dots, A^{-N}B \right ] \times u^{(i)}_{0, N-1}(x_{0})
 \end{align}
 Then, corresponding to the above 4 conditions, the dismension of the solution space $U^{(i)}_{N}(x_{0})$ of the Eq. \eqref{eq:a0236} are as follows
  \begin{align} 
 (a1) \;
 &  \dim U^{(1)}_{N}(x_{0})  = \dim U^{(2)}_{N}(x_{0}) =N-k, \;\;k=1,k_1
 \label{eq:a0237a} \\
 (a2) \;
 &  \dim U^{(1)}_{N}(x_{0})   =N-k<N-k+1= \dim U^{(2)}_{N}(x_{0}), \;\;k=1,k_1
 \label{eq:a0237b} \\
( a3) \;
 &  \dim U^{(1)}_{N}(x_{0})  = \dim U^{(2)}_{N}(x_{0}) =N-k+1, \;\;k=1,k_1
 \label{eq:a0237c} \\
 (b1) \;
 &  \dim U^{(1)}_{N}(x_{0})   =N-k_1<N-k_2= \dim U^{(2)}_{N}(x_{0})
 \label{eq:a0237d} \\
 (b2) \;
 &  \dim U^{(1)}_{N}(x_{0})   =N-k_1<N-k_2+1= \dim U^{(2)}_{N}(x_{0})
 \label{eq:a0237e} \\
 (b3) \;
 &  \dim U^{(1)}_{N}(x_{0})   =N-k_1+1 <N-k_2+1 = \dim U^{(2)}_{N}(x_{0})
 \label{eq:a0237f} 
  \end{align}
Furthermore, if we have 
	 \begin{align} \label{eq:a0228b}
R^{(1)} (i) \subset R^{(2)} (i), \; \forall i \le N, 
\end{align}
then the given state $x_0$ and the corrsponding solution space $U^{(i)}_{N}(x_{0})$ are satisfied only cases a2), a3), b2), and b3) in above 6 cases. 
 Hence, whether the two controllability regions satisfy Eq. \eqref{eq:a0228} or \eqref{eq:a0228b}, we have
 \begin{align} \label{eq:a0241}
 \dim U^{(1)}_{N}(x_{0}) \le \dim U^{(2)}_{N}(x_{0})
 \end{align}
So, considered that the higher the space dimension is and the more the number of the states in the state space is, for any state $x_{0} \in R^{(2)} (N)$, 
 the solution space of the system $ \Sigma_2$ is larger than that of the system $ \Sigma_1$, and then the systems $ \Sigma _2$ for controlling the state to $x_0$ to the origin will be with more control strategies than the system $ \Sigma _1$ .
 \qed

By { \bfseries Theorem \ref{th:t0301} }, we have the following discussions:

(1) Not only the time waste can be reduced by promoting the control ability, but also other closed-loop performance related the control time waste can be improved.

(2) In fact, that the solution space of the input variables for the control problems is bigger implies that the control strategies in the solution space are with better robustness, and then the closed-loop control systems is also with better robustness. 

Therefore, optimizing the open-loop control ability are with very greater signification for these practical control engineering problems and it's very necessary to optimize the control ability. To optimize the control ability, it is necessary to establish the quantify analysis and computing method for the control ability. Paper \cite{zhaomw202001} prove an analytical computing equation for the volume of the controllability region and deconstruct the volume equation to construct some analytical factors about the shape of the controllability region. Based on these analytical expressions of the volume and shape factors, the optimizing and promoting methods for the control ability can be set up conveniently.

The analytic expressions of the volume and shape factors of the controllability region, which can be regard as a special zonotope generated by the matrix pair, is got in paper \cite {zhaomw202001} and \cite {zhaomw202004} . Based on the analytic computation of the volume and shape factors, comparing the size of the controllability regions of LDT systems is become possible, and then by \textbf{Theorem \ref{th:t0301}}, we can compare the control ability between the difference LDT systems. Furthmore, according the control ability computing, we can choose the controlled plant for constructing engineering equipment systems, determine the input variables for the feedback control systems, place the location of the actuator for the lager mechanical systems, etc. 

 \section {Numerical Experiments}

 \begin{example}
	Considered the following LDT models $ \Sigma_d \left(A_d, B_d \right)$ and $ \Sigma_a \left(A_a, B_a \right)$ for DC motor with the excitation controller and AC motor with the variable frequency controller, respectively
 \begin{align}
 & \;
x_{d}(k+1)= \left[ \begin{array}{ccc}
0 & 1 & 0 \\
0 & 0 & 1 \\
0.69527 & -2.3565 & 2.660
 \end{array} \right] x_{d}(k)+ \left[ \begin{array}{c}
0 \\ 0 \\
8.74
 \end{array} \right]u_{d}(k) \notag \\
& \;
x_{a}(k+1)= \left[ \begin{array}{ccc}
0 & 1 & 0 \\
0 & 0 & 1 \\
0.65711 & -2.2691 & 2.610
 \end{array} \right]x_{a}(k)+ \left[ \begin{array}{c}
0 \\ 0 \\
19.17
 \end{array} \right] u_{a}(k) \notag 
 \end{align}
where the input variable $u_{d}$ and $u_{a}$ are the input voltages, and the state variable $x_{d}$ and $x_{a}$ are consist of the main circuit current, the speed and acceleration of two motors, respectively. The rated values of the input variable and state variables are respectively as
 \begin{align}
 \left( u_{d}^*, i_{d}^*, y_{d}^*, \dot y_{d}^* \right) =( \textnormal{24V, 30A, 200rad/s, 30rad/s}^2) \notag \\
 \left( u_{a}^*, i_{a}^*, y_{a}^*, \dot y_{a}^* \right) =( \textnormal{12V, 30A, 230rad/s, 35rad/s}^2) \notag
 \end{align}
, these state variables are expected in operation intervals with the upper bounds \textnormal { \{30A, 180rad/s, 30rad/s}$^2$ \}.

The problem here is which motor is with the stronger control ability in the rater interval and expecting interval, respectively.

 \end{example}

The computing and analyzing process are introducted as follows.

(1) By Eqs. \eqref{eq:a0206} and \eqref{eq:a0207}, for comparing the control ability of the two motors in its variable rated intervals, the system models $ \Sigma \left( A_d, B_d \right)$ and $ \Sigma \left( A_a, B_a \right)$ are transformed respectively as 
 \begin{align}
& 	 \left( A_d, b_d \right)= \left( \left[ \begin{array}{ccc}
 0 & 6.6667 & 0 \\
0 & 0 & 0.1500 \\
0.6953 & -15.7100 & 2.6600
 \end{array} \right], \;
 \left[ \begin{array}{c}
0 \\
0 \\
6.9920
 \end{array} \right] \right) \notag \\
&	 \left( A_a, b_a \right)= \left( \left[ \begin{array}{ccc}
 0 & 7.6667 & 0 \\
0 & 0 & 0.1522 \\
0.5632 & -14.9112 & 2.6100
 \end{array} \right], \;
 \left[ \begin{array}{c}
0 \\
0 \\
6.5726
 \end{array} \right] \right) \notag
 \end{align}

\begin{figure}[htbp]
	 \centering
	 \begin{minipage}[c]{0.9 \textwidth} 
		 \centering
		 \includegraphics[width=0.8 \textwidth]{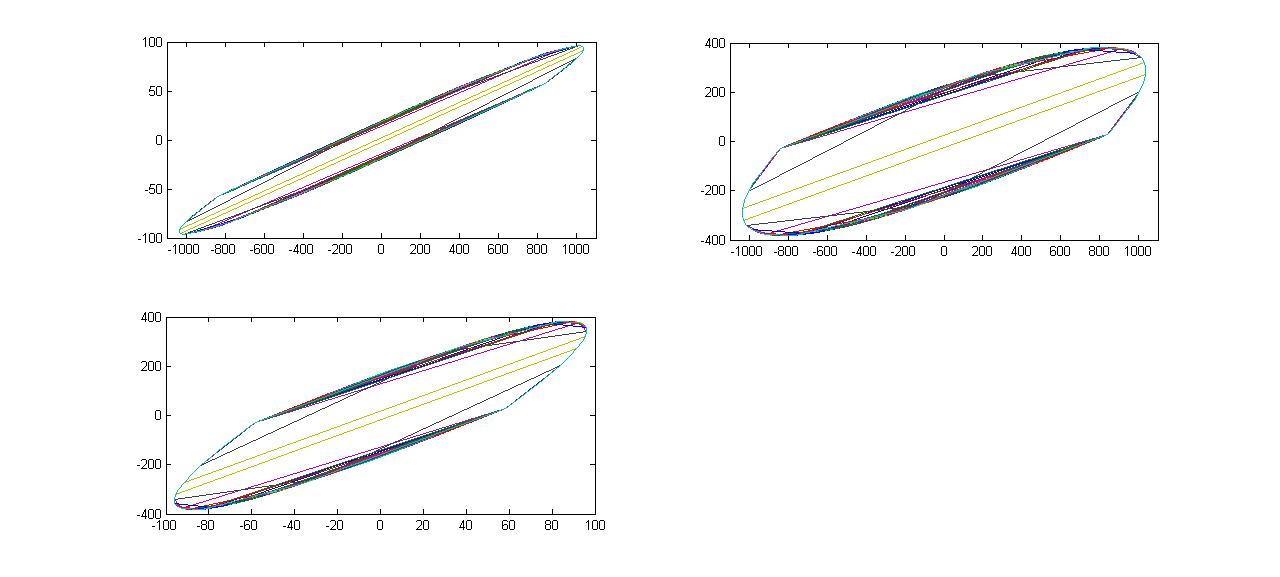} 
		 \\ (a) \footnotesize {DC motor in the rated intervals with the 2-D shape factors \{0.1920, 0.4134, 0.2405 \}}
	 \end{minipage}%
 \\
	 \begin{minipage}[c]{0.9 \textwidth}
		 \centering
		 \includegraphics[width=0.8 \textwidth]{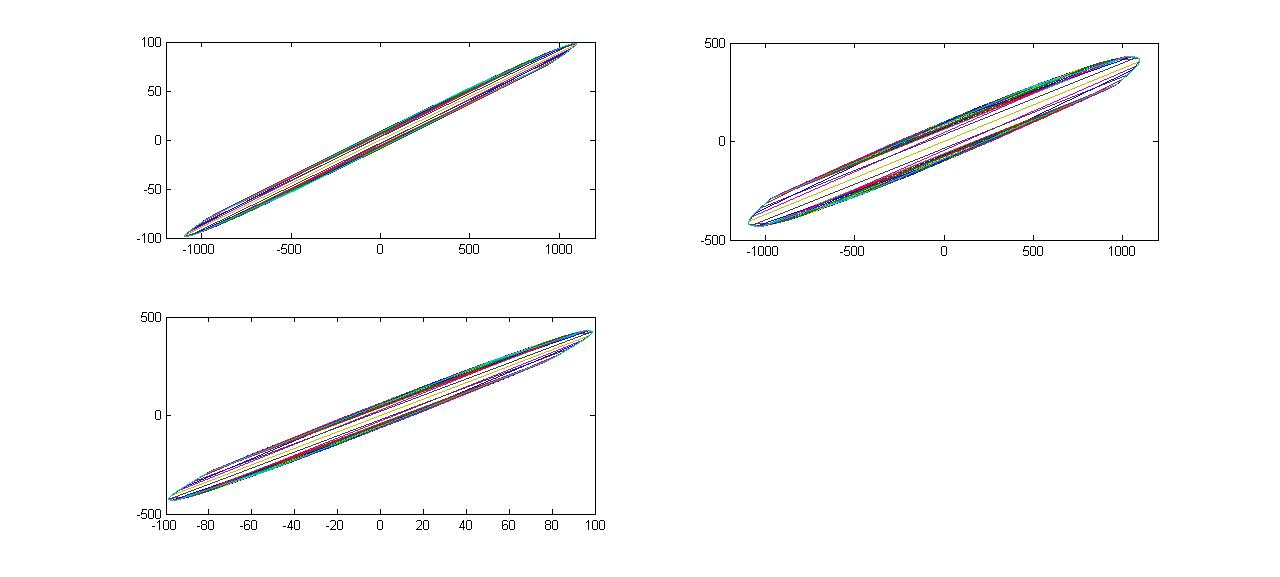}
		 \\ (b) \footnotesize {AC motor in the rated intervals with the 2-D shape factors \{0.1474, 0.3262, 0.1878 \}}
	 \end{minipage}
	 \caption[c]{The 3-D controllability region with the 2-D shape factors $ \left \{f_{1, 1, 2}, f_{1, 1, 3}, f_{1, 2, 3} \right \}$ illustrated by 3 2-D projection drawing \label {fig:0302}}	
 \end{figure}

The computing results for the volumes, the shape factors $ \left \{f_1, f_{1, 1, 2}, f_{1, 1, 3}, f_{1, 2, 3} \right \} $, and the side lengths $ \left \{f_{2, 1, 2}, f_{2, 1, 3}, f_{2, 2, 3} \right \}$ of the circumscribed hypercube (or circumscribed rhomboid) of the controllability regions \cite{zhaomw202004} are shown in \textbf{Table \ref{tab:t0301}}, and \textbf{Fig. \ref {fig:0302}} illustrate the 3-D controllability regions by the 3 2-D projection drawing in $x_1-x_2, \; x_1-x_3$, and $x_2-x_3$ planes. By the figures, we can see, the controllability region of the AC motor is more flatted than that of the DC motor, and accordingly its shape factor $f_1$ is less than the DC motor.

By paper \cite{zhaomw202004}, the size of the controllability regions are depend on its
volume and shape described by the shape factors and the side lengths. When the side lengths are approximated, that is, the circumscribed rhomboids of the controllability regions are approximated, the greater the value of the factor $f_1$ is, the bigger the volume of the region is, and then the bigger the region is. When the value of the factor $f_1$ are approximated, the greater the side lengths are, and then the bigger the region is.
By \textbf{Table \ref{tab:t0301}} and \textbf{Fig. \ref {fig:0302}}, the controllability region of the DC motor is bigger than the AC motor in the rated intervals, and then, based on \textbf{Definitions \ref {de:d0301} and \ref {de:d0302}}, the controll ability of the DC motor in the rated intervals is stronger than the AC motor. Therefore, by \textbf{Theorem \ref {th:t0301} }, the excitation controller for the DC motor is with the bigger parameter space designing the controller or the control law, and then the controller with the better closed-loop time-attribute performance index and robustness can be gotten than the variable frequency controller
for the AC motor.

\begin{table}
	\centering
	\caption{Numerical results for the reachable regions}
	\label{tab:t0301} 
	\begin{tabular}{ccccc}
		\hline \noalign{ \smallskip}
		& \multicolumn{2}{c}{ \textbf{rated interval }} & \multicolumn{2}{c}{ \textbf{expectivng interval }} \\
		factors & DC & AC & DC & AC \\
		\hline
		\noalign{ \smallskip}
		volume & 3.5038e7 & 1.4307e7 & 3.8931e7 & 2.1328e7 \\
		$f_1$ & 0.0191 & 0.0090 & 0.0191 & 0.0090 \\
		$f_{1, 1, 2}$ & 0.1920 & 0.1474 & 0.1920 & 0.1474 \\
		$f_{1, 1, 3}$ & 0.4134 & 0.3262 & 0.4134 & 0.3262 \\
		$f_{1, 2, 3}$ & 0.2405 & 0.1878 & 0.2405 & 0.1878 \\
		$f_{2, 1}$ & 1.2330e4 & 1.6035e4 & 1.2345e4 & 1.6801e4 \\
		$f_{2, 2}$ & 4.0026e4 & 4.4408e4 & 4.0077e4 & 4.6827e4 \\
		$f_{2, 3}$ & 3.5779e4 & 3.3471e4 & 3.5825e4 & 3.5511e4 \\
		\noalign{ \smallskip} \hline
	\end{tabular}

\end{table}

(2) By Eqs. \eqref{eq:a0209} and \eqref{eq:a0210}, for comparing the control ability of the two motors in the expecting intervals for these state variables, the system models $ \Sigma \left( A_d, B_d \right)$ and $ \Sigma \left( A_a, B_a \right)$ are transformed respectively as 
 \begin{align}
& 	 \left( A_d, b_d \right)= \left( \left[ \begin{array}{ccc}
 0 & 6.0000 & 0 \\
0 & 0 & 0.1667 \\
0.6953 & -14.1390 & 2.6600
 \end{array} \right], \;
 \left[ \begin{array}{c}
0 \\ 0 \\
 6.9920
 \end{array} \right] \right) \notag \\
&	 \left( A_a, b_a \right)= \left( \left[ \begin{array}{ccc}
 0 & 6.0000 & 0 \\
0 & 0 & 0.1667 \\
0.6571 & -13.6146 & 2.6100
 \end{array} \right], \;
 \left[ \begin{array}{c}
0 \\ 0 \\
7.6680
 \end{array} \right] \right) \notag
 \end{align}

\begin{figure}[htbp]
	 \centering
	 \begin{minipage}[c]{0.9 \textwidth} 
		 \centering
		 \includegraphics[width=0.8 \textwidth, height=0.5 \textwidth]{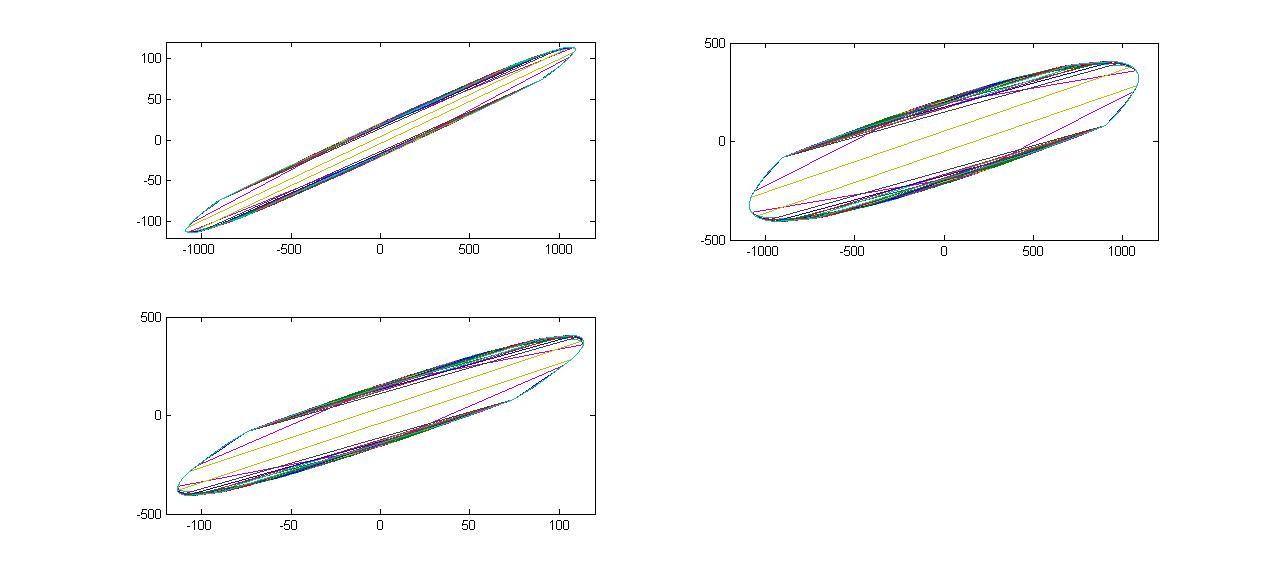} 
		 \\ (a) \footnotesize {DC motor in the expecting interval}
	 \end{minipage}	
	 \\
	 \begin{minipage}[c]{0.9 \textwidth}
		 \centering
		 \includegraphics[width=0.8 \textwidth, height=0.5 \textwidth]{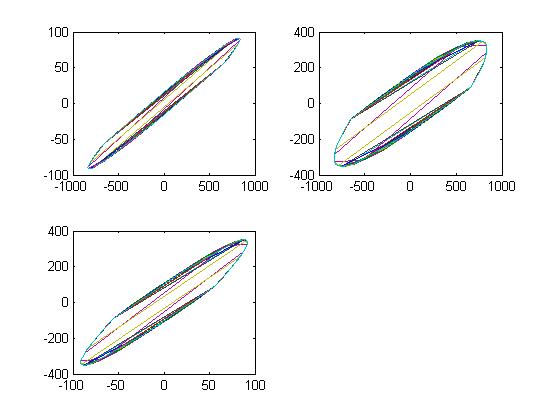}
		 \\ (b) \footnotesize {AC motor in the expecting interval}
	 \end{minipage}
	 \caption[c]{The 3-D controllability region illustrated by 3 2-D projection drawing \label {fig:0303}}	
 \end{figure}

The computing results are shown in \textbf{Table \ref{tab:t0301}}, and the 3-D controllability regions are illustrated by the 3 2-D projection drawing in \textbf{Fig. \ref {fig:0303}}. By the figures, we can see, the controllability region of the AC motor is more flatted than that of the DC motor, and accordingly its shape factor $f_1$ is less than the DC motor.

 From \textbf{Table \ref{tab:t0301}} and \textbf{Fig. \ref {fig:0303}}, we can see, the controllability region of the DC motor is bigger than the AC motor in the expecting variable intervals, and then, based on \textbf{Definitions \ref {de:d0301} and \ref {de:d0302}}, the controll ability of the DC motor in the rated intervals is stronger than the AC motor. Therefor, by \textbf{Theorem \ref {th:t0301} }, the better controller for the DC motor can be designed than the AC motor.

 \section {Conclusions}

		In this article, the definition on the time-attribute control ability, and the relation between the open-loop control ability and the closed-loop performance are studied systematically. Firstly, to define and compare the state control ability, the normalization of the input variables and state variables in the different control plants or one controlled plant with the different system parameters are discussed. With the help of the normalization, the time-attribute control ability with the unit input constraint (input amplitude limited) can be defined. Finally, a theorem on the relations among the open-loop control ability, the control strategy space (i.e., the solution space of the input variables for control problems), and the closed-loop time performance is purposed and proven. Therefore, it is necessary to optimize the control ability for the practical engineering problems. Based on the results in paper \cite{zhaomw202001}, the optimizing and promoting methods for the control ability can be set up conveniently, and then the controller with the greater designing parameter space and then the better closed-loop peroformance index and robustens can be got. 
		
	 \bibliographystyle{model1b-num-names}
 \bibliography{zzz}

 \end{document}